\documentclass[10pt]{iopart}
\usepackage{psfig}
\begin{document}

\title[Strange quark matter: 
mapping QCD lattice results to finite baryon density]
{Strange quark matter: 
mapping QCD lattice results to finite baryon density
by a quasi-particle model\footnote[1]{Supported by BMBF 06DR921 and GSI.}}

\author{B K\"ampfer*, A Peshier\P\footnote[6]{Supported by 
A.\ v.\ Humboldt foundation.}  
and G Soff\S}

\address{*\
Forschungszentrum Rossendorf, D-01314 Dresden, PF 510119, Germany}
\address{\P\
Dept.\ of Physics, Brookhaven Nat.\ Lab., Upton, NY 11973, USA}
\address{\S\
Institut f\"ur Theoretische Physik, TU Dresden, D-01062 Dresden, Germany}

\begin{abstract}
A quasi-particle model is presented which describes QCD lattice results
for the 0, 2 and 4 quark-flavor equation of state. 
The results are mapped to finite baryo-chemical potentials. 
As an application of the 
model we make a prediction of deconfined matter with appropriate
inclusion of strange quarks and consider pure quark stars.
\end{abstract}

\vspace*{-3mm}

\section{Introduction}

The QCD lattice calculations of the equation of state (EoS)
of deconfined
matter have advanced to such a level that reliable results for the pure
gluon plasma are available \cite{N_f=0}.
For the two-flavor case an estimate of the continuum extrapolation
is at disposal \cite{N_f=2}. The EoS of four light flavors
\cite{N_f=4} are not yet continuum extrapolated.
The physically interesting
case of two light flavors and a medium-heavy strange quark is still in
progress \cite{N_f=3}. All of these {\it ab initio} calculations of the EoS
of deconfined matter, however, are yet constrained
to finite temperatures $T \ge T_c$ (here $T_c$ is the deconfinement temperature)
and vanishing baryo-chemical potential, $\mu = 0$. While first attempts
to calculate the EoS at $\mu \ne 0$ are under way \cite{finite_mu},
the final results can probably be expected only in a few years.
It is therefore a challenge to attempt an extrapolation of the QCD lattice
results into the domain of finite baryo-chemical potential. Here we employ
a quasi-particle model to accomplish this goal.

Quasi-particle models have proven powerful in describing properties
of strongly correlated systems in condensed matter physics.
Analogously, one should expect that also strongly interacting matter
can be described effectively within quasi-particle models.
Indeed, the investigations of $\phi$-derivable self-consistent
approximations \cite{Baym}, combined with hard thermal loop resummation
\cite{Braaten}, delivered recently compelling support of 
a quasi-particle description of deconfined matter. Starting from the QCD
Lagrangian a chain of approximations is derived \cite{Blaizot}
which results in a
quasi-particle model of deconfined matter agreeing with lattice results
at $T > 2.5 T_c$. On the other hand, employing the hard thermal/dense
loop resummation at finite baryo-chemical potential, further
evidence for a quasi-particle description of cold deconfined matter
is gained \cite{Baier_Redlich}. What is still needed is an interpolating
model, which reproduces the QCD lattice data down to $T_c$ and, at the
same time, extrapolates to finite values of $\mu$ even up to
$T = 0$. We present here such a model and apply it to calculate static
properties of cold, pure quark stars with strange quarks
properly taken into account.

\section{Quasi-particle model of deconfined matter}

With increasing sophistication of QCD lattice calculations of the EoS
also phenomenological quasi-particle models have been developed
\cite{quasiparticle_models}. Of central importance to our 
model \cite{PRC2000} are the baryon density $n$ and the
entropy density $s$ as quantities which are dominated by the 
quasi-particle structure of the interacting system.\footnote{
This is supported by the $\phi$-derivable
approach \protect\cite{Baym}
where corrections to the quasi-particle picture arise only beyond the
resummed leading-loop order contributions 
\protect\cite{Blaizot,Baym_2}. 
Within the massless $\varphi^4$ theory
the form of $s$ below is obtained
by employing the Luttinger-Ward theorem \protect\cite{Luttinger_Ward}
with a super-daisy resummed propagator and a double-ring  
$\phi$ functional \protect\cite{phi}.}
Approximating the self-energies of the quarks ($q$), with a current mass
$m_{q 0}$,
and the gluons ($g$) by the gauge-invariant asymptotic values of the 1-loop
expressions
\begin{eqnarray}
  \Pi_q^*
  & = &
  2 \, \omega_q \, (m_{q 0} + \omega_q) \, ,
  \quad
  \omega_q^2
  =
  \frac 16 \, 
  \left[ T^2 + \frac{\mu_q^2}{\pi^2} \right]
  g_{\rm eff}^2, \\
  \Pi_g^*
  &=&
  \frac16
  \left[
     \left( 3 + \frac 12 \, N_f \right) T^2
   + \frac{3}{2\pi^2} \sum_q \mu_q^2
  \right] g^2_{\rm eff},
 \label{Pi}
 \end{eqnarray}
the densities are given by the standard formulae of ideal gases
(labeled by the superscript ''id'') of quarks and
gluons with effective masses $m_i^2 = m_{i 0}^2 + \Pi_i^*$, $i = (q,g)$,
\begin{eqnarray}
n & = & \sum_q \left\{ n_q^{\rm id} (T, \mu; m_q[T,\mu]) -
       n_{\bar q}^{\rm id} (T, \mu; m_q[T,\mu]) \right\}, \\
s & = & s_g^{\rm id}(T, \mu; m_g[T,\mu]) +
\sum_q s_q^{\rm id}(T, \mu; m_q[T,\mu]).
\label{entropy}
\end{eqnarray}
Beyond this resummation of the leading-order contributions, non-perturbative
effects are described in the phenomenological quasi-particle 
model by the effective coupling $g_{\rm eff}$. 
The requirement
$g_{\rm eff}(T, \mu) \to g_{\rm 1-loop}$ at large values of $T$ and/or $\mu$
ensures the smooth transition to the asymptotic regime.

The corresponding pressure $p$ and energy density $e$ are
$p = \sum_i p^{\rm id}_i ( T, \mu; m_i[T, \mu]) - B(T, \mu)$ and
$e  =  \sum_i e^{\rm id}_i ( T, \mu; m_i[T, \mu]) + B(T, \mu)$.
The quantity $B(T, \mu)$ is not an independent quantity but
obtained by integrating
$(\partial p^{\rm id} / \partial m^2_i)$ 
$(\partial \Pi^*_i / \partial T) 
=
(\partial B / \partial T)$,
$(\partial p^{\rm id} / \partial m^2_i) 
(\partial \Pi^*_i / \partial \mu)
=
(\partial B / \partial \mu )$ 
which come from the stationarity condition
$\delta p / \delta \Pi_i = 0$
\cite{Lee_Yan}.

Let us mention two implications of the quasi-particle model.
(i) According to the Feynman-Hellmann relation the chiral condensate 
is given by
$\langle \bar \Psi_q \Psi_q \rangle \propto 
\partial p / \partial m_{q 0} \to 0$ for
$m_{q 0} \to 0$,
i.e. for vanishing current quark masses 
the chiral condensate vanishes
in agreement with the chiral symmetry restoration at $T > T_c$.
(ii) In the asymptotic region, $T \to \infty$ and $\mu \to \infty$,
an expansion in the coupling yields
$p(T, \mu) = p_0(T, \mu) + p_2(T, \mu) + \cdots$ thus
reproducing the perturbative results \cite{perturbative_expansion}
in the orders of $g^0$ and $g^2$. 

\section{Tests of the model \label{tests}}

\begin{figure}[t]
 \centerline{
 \psfig{file=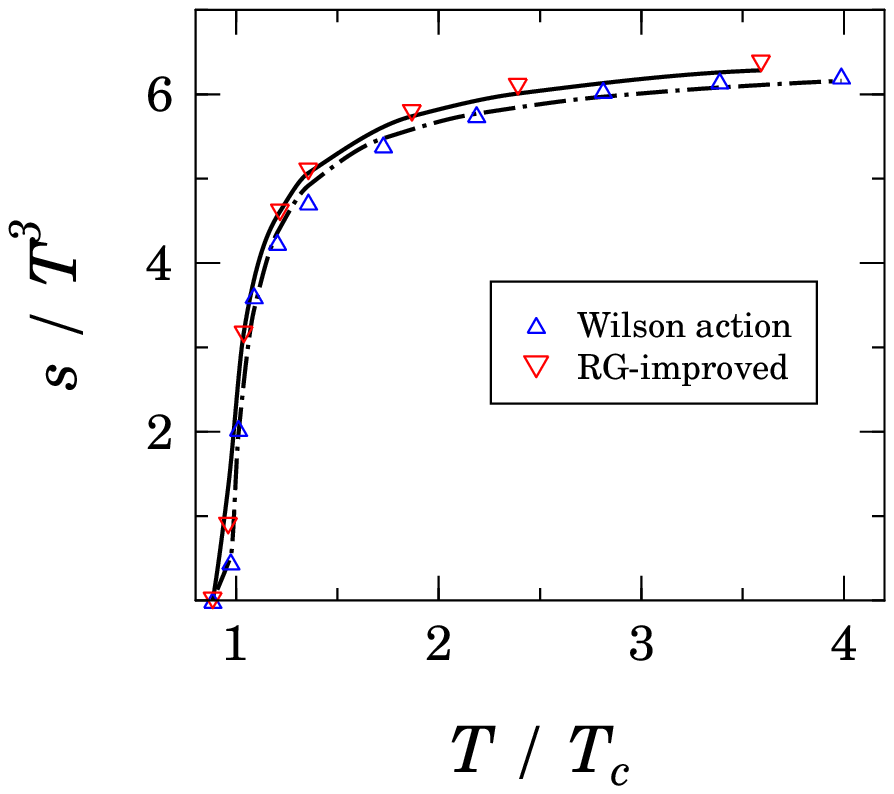,width=4.4cm,angle=-0} \hfill
 \psfig{file=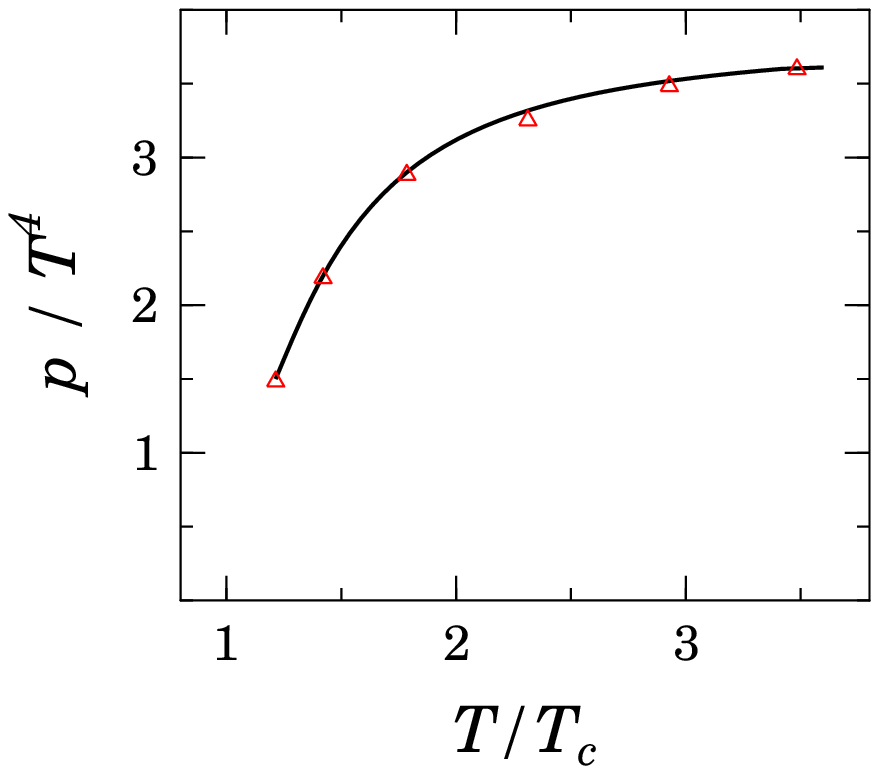,width=4.4cm,angle=-0} \hfill
 \psfig{file=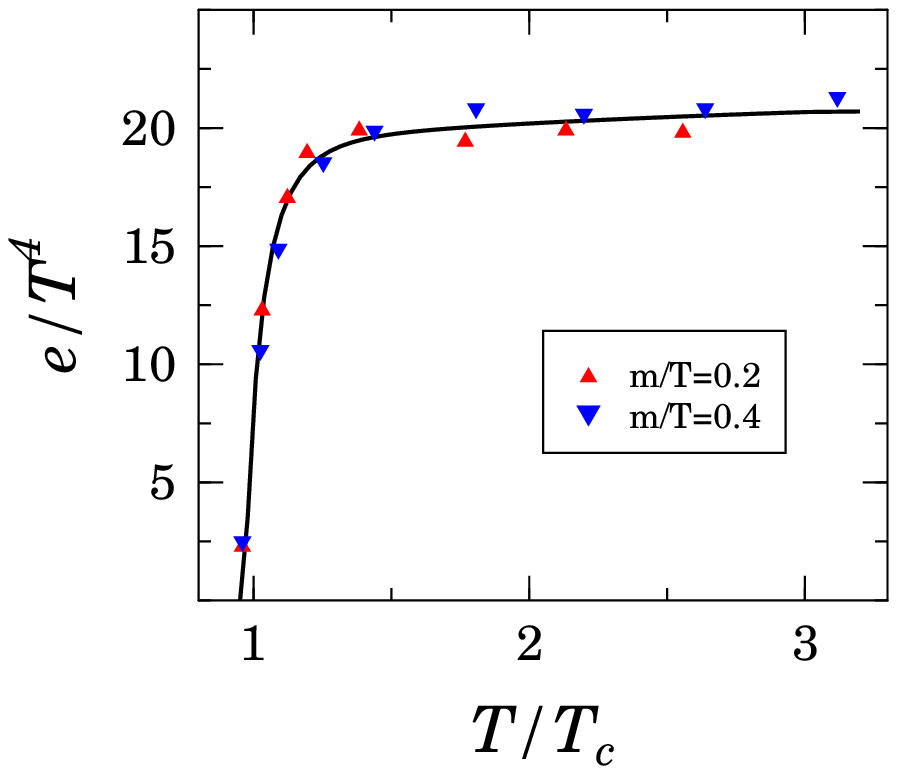,width=4.4cm,angle=-0}}
\caption{
Comparison of the quasi-particle model with QCD lattice results
for the case of $N_f = 0$ (left, \cite{N_f=0}), 
$N_f = 2$ (middle, \cite{N_f=2}) and
$N_f = 4$ (right, \cite{N_f=4}).}
\label{comparisons}
\end{figure}

For the effective coupling strength $g_{\rm eff}$
we chose a regularized parameterization
of the 1-loop running coupling strength.
The resulting comparison with the QCD lattice data is displayed in 
figure \ref{comparisons} for various flavor numbers $N_f$.
Notice the almost perfect agreement with the data which ensures that the
other thermodynamical quantities are also nicely reproduced
(for details cf.\ \cite{PRC2000}).

\section{Extrapolation to $\mu > 0$}

Our model assumes 
the validity of the underlying
quasi-particle structure at finite $\mu$. Some limitation
of this assumption may arise from the recently debated
supra-fluidity and color-flavor locking effects at small temperatures
and asymptotically large chemical potentials \cite{no_deconfinement}.
However, since many of the discussed effects \cite{no_deconfinement},
which also depend sensitively on the actual strange quark mass
(as the phase diagram does already at $\mu = 0$ \cite{Karsch_phase_diagram}),
are near the Fermi surface, the gross properties of the EoS might not 
be modified strongly.

Since the pressure $p$ is a potential it has to fulfill 
the Maxwell relation
$\partial n / \partial T = \partial s / \partial \mu$,
which results in a partial differential equation for the coupling 
of the form
\begin{equation}
a_T\, \frac{\partial g_{\rm eff}^2}{\partial T}
+ a_\mu\, \frac{\partial g_{\rm eff}^2}{\partial \mu}
= b.
\label{charac_eq.}
\end{equation}
\begin{figure}[b]
~\vskip -5mm
\centerline{
\psfig{file=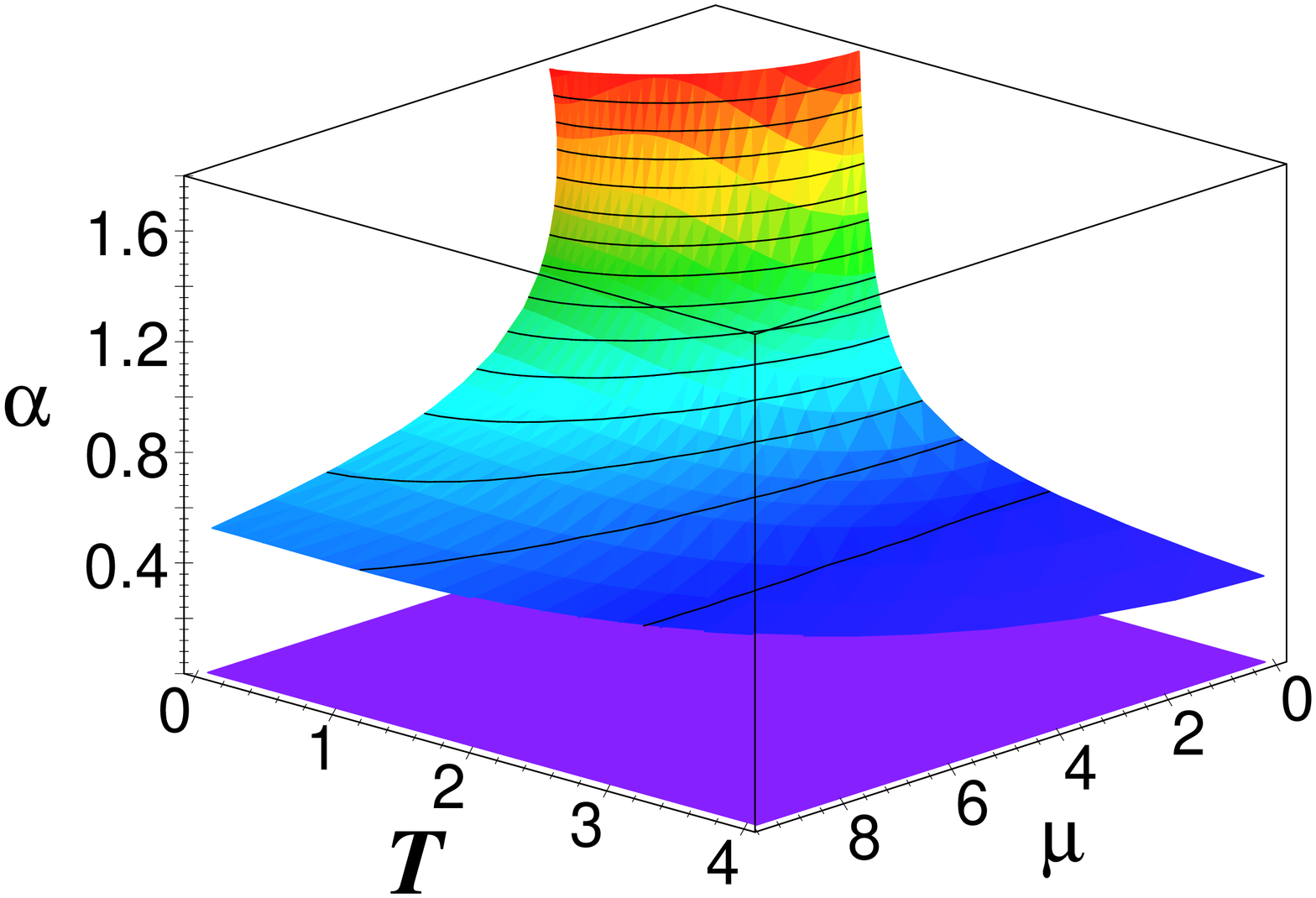,width=6cm,angle=-90} \hspace*{0.3cm}
\psfig{file=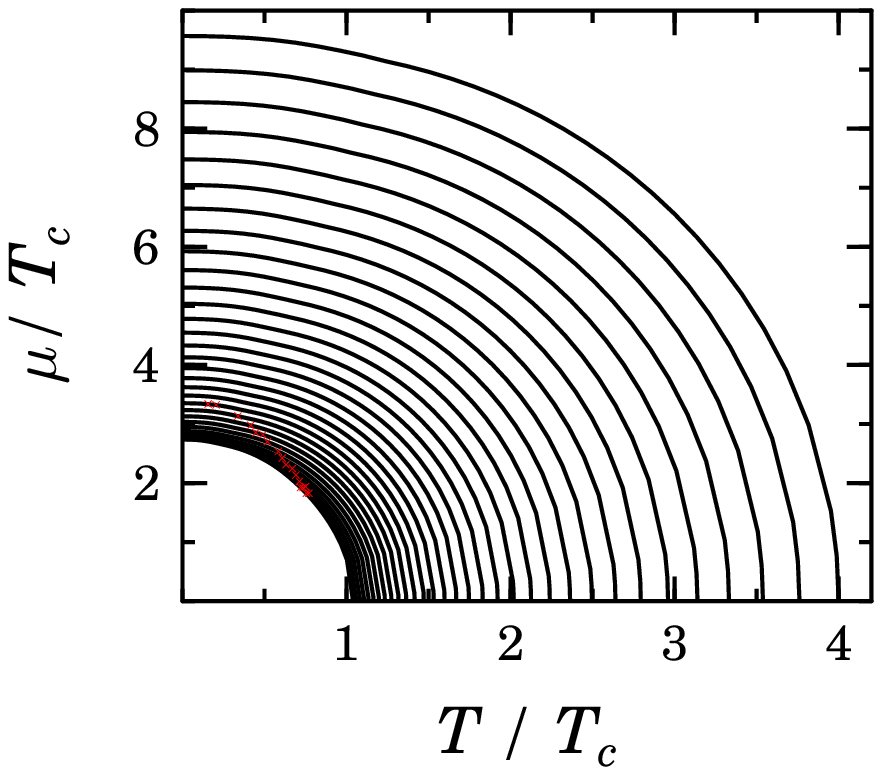,width=6cm}}
\caption{Left panel:
The effective coupling strength 
$\alpha = g_{\rm eff}^2/(4 \pi)$
as a function of $T$ and $\mu$ (both in units of $T_c$)
Right panel: 
The characteristics in solving equation (\ref{charac_eq.}).}
\label{characteristics}
\end{figure}
A solution for the 4-flavor lattice case \cite{N_f=4}
is displayed in the left panel of figure \ref{characteristics}.
The strong increase of $g_{\rm eff}$ towards smaller values of $T$ and
$\mu$ can be considered as an indication of approaching the phase boundary. 
Despite of the fact that our model does not allow to make statements
on the nature of the deconfinement transition, one can safely conclude
that the region of smaller values of $T$ and $\mu$, not covered
by the elliptic characteristics shown in the right panel of 
figure \ref{characteristics}, 
may be related to confined matter.
The order of magnitude of the baryo-chemical potential
$\mu_B = 3 \mu$ for the innermost characteristics
at $T = 0$ is 1.4 GeV when choosing the scale
$T_c = 160$ MeV. 

Thermodynamic state variables
are available in tabular form in \cite{Peshier}.
At small $T$, the pressure in the small-$\mu$ region becomes negative
giving a lower bound for the existence of conventional deconfined matter.
The energy per baryon,
however, at the point of vanishing pressure is quite large,
typically in the order of 1.5 GeV for $N_f = 2, 3 ,4$.
That means, even when including strangeness,
no hint to stable strangelets arises from our model.

\section{Pure quark stars with strangeness}

To condense the properties of the predicted EoS in a few concise
numbers let us consider pure quark stars with locally charge-neutral
deconfined matter in $\beta$ equilibrium. The chemical potentials
of u, d, s quarks and leptons ($l$)
are related according to $\mu_s = \mu_d = \mu_u + \mu_l \equiv \mu$.
Unfortunately, continuum extrapolated QCD lattice data for the 2 + 1
flavor case with physical quark masses are not yet at disposal
at $T \ne 0$.
Therefore, we first guess sets of EoS for the 2 + 1 flavor case
at $\mu = 0$ and then extrapolate them to finite $\mu$ at $T = 0$.
In doing so we are guided by the parameter range obtained in \cite{PRC2000}.
Adding to the pressure the lepton contribution it turns out that the
energy density - pressure relation obtained numerically can be
approximated by $e = 4 \tilde B + \tilde \alpha p$ with
$\tilde \alpha = 3.1 \cdots 4.5$ and
$\tilde B^{1/4} \ge 200$ MeV. Note that $\tilde B^{1/4}$ is not
longer an {\it ad hoc} parameter in the spirit of the bag model
but a parameter which will be fixed by improving the QCD lattice
data for the 2 + 1 flavor case \cite{N_f=3}.
\begin{figure}[b]
\centerline{
\psfig{file=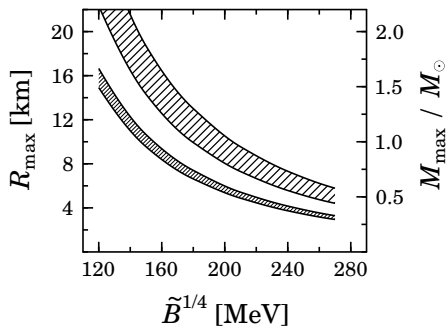,width=6.5cm}}
\caption{
Maximum masses (upper hatched band,
right scale) and corresponding radii (lower hatched band, left scale)
of pure quark stars
as a function of $\tilde B^{1/4}$.
The upper (lower) limit of the bands is for $\tilde \alpha = 3$ (4.5).}
\label{q_stars}
\end{figure}

Integrating the Tolman-Oppenheimer-Volkoff equations for
spherical stars in hydrodynamical equilibrium up to
vanishing pressure at the surface one gets the masses and radii
as displayed in figure \ref{q_stars}. 
Since the estimated range
of $\tilde B^{1/4}$ is $> 200$ MeV, the strange quark stars
have small masses ($M \le 1 M_\odot$) and radii ($R \le 7$ km).

It should be emphasized that our model does not predict absolutely
stable (strange) quark matter. Rather, at smaller values of $\mu$ and
$p > 0$ the transition to hadron matter is expected to happen.
The static properties of the corresponding hybrid stars 
depend sensitively on the nature of the confinement
transition and details of the hadronic EoS. 

\section{Summary}

In summary we present a quasi-particle model for the EoS of purely deconfined
matter, which describes the available QCD lattice calculations
at $\mu = 0$ and $T \ge T_c$, and map this to finite values of the
baryo-chemical potential including properly the strangeness degree of
freedom. The model does not point to stable strangelets.
The resulting cold, pure quark stars with strangeness are light and small.

At finite temperatures the strange quarks do not play any exceptional
role: the self-energies mask nearly all of the finite-mass effect,
even having used here $m_{s 0} = 150$ MeV, while recent
lattice calculations point to smaller values
\cite{Wittig}.
This is in agreement with QCD lattice calculations which display
for $p/p_0$ almost the same values for the 2, 3 and 2 + 1 flavor
cases \cite{N_f=3}. 
The gluon abundance near $T_c$ is strongly
suppressed.

\end{document}